# Unexpectedly High Cross-plane Thermoelectric Performance of Layered Carbon Nitrides

Zhidong Ding[1,#], Meng An[1,2,#], Shenqiu Mo[1,3], Xiaoxiang Yu[1,3], Zelin Jin[1,3], Yuxuan Liao[4], Jing-Tao Lü[5,*], Keivan Esfarjani[6], Junichiro Shiomi[4,7,*], Nuo Yang[1,3,*]

[1] State Key Laboratory of Coal Combustion, Huazhong University of Science and Technology, Wuhan 430074, China

[2] College of Mechanical and Electrical Engineering, Shaanxi University of Science and Technology, Xi'an, 710021, China

[3] Nano Interface Center for Energy (NICE), School of Energy and Power Engineering, Huazhong University of Science and Technology, Wuhan 430074, China

[4] Department of Mechanical Engineering, The University of Tokyo, 7-3-1 Hongo, Bunkyo, Tokyo 113-8656, Japan

[5] School of Physical and Wuhan National High Magnetic Field Center, Huazhong University of Science and Technology, Wuhan 430074, China

[6] Department of Mechanical and Aerospace Engineering, University of Virginia, Charlottesville 22904, United States

[7] Center for Materials Research by Information Integration, National Institute for Materials Science, 1-2-1 Sengen, Tsukuba, Ibaraki 305-0047, Japan

# Z.D. and M.A. contributed equally to this work.
* To whom correspondence should be addressed. E-mail: (J.-T.L.) jtlu@hust.edu.cn, (J.S.) shiomi@photon.t.u-tokyo.ac.jp, (N.Y.) nuo@hust.edu.cn

**Abstract:** Organic thermoelectric (TE) materials create a brand new perspective to search for high-efficiency TE materials, due to their small thermal conductivity. The overlap of $p_z$ orbitals, commonly existing in organic π-stacking semiconductors, can potentially result in high electronic mobility comparable to inorganic electronics. Here we propose a strategy to utilize the overlap of $p_z$ orbitals to increase the TE efficiency of layered polymeric carbon nitride (PCN). Through first-principles calculations and classical molecular dynamics simulations, we find that A-A stacked PCN has unexpectedly high cross-plane *ZT* up to 0.52 at 300 K, which can contribute to n-type TE groups. The high *ZT* originates from its one-dimensional charge transport and small thermal conductivity. The thermal contribution of the overlap of $p_z$ orbitals is investigated, which noticeably enhances the thermal transport when compared with the thermal conductivity without considering the overlap effect. For a better understanding of its TE advantages, we find that the low-dimensional charge transport results from strong $p_z$-overlap interactions and the in-plane electronic confinement, by comparing π-stacking carbon nitride derivatives and graphite. This study can provide a guidance to search for high cross-plane TE performance in layered materials.

**Introduction**

Thermoelectric (TE) technology can directly convert heat into electricity as well as being used in solid-state cooling. Thermoelectric devices are considered for a variety of applications [1–5], for example, distributed power generation, waste heat recovery from vehicles, and power supply to electronic devices. The efficiency of TE conversion can be measured by a dimensionless figure of merit $ZT$ that is defined as $ZT = \sigma S^2 T/(\kappa_e + \kappa_{ph})$. $\sigma$ is the electrical conductivity; $S$ is the Seebeck coefficient; $T$ is the absolute temperature; $\kappa_e$ is the electronic thermal conductivity; $\kappa_{ph}$ is the lattice thermal conductivity. Reducing $\kappa_{ph}$ while maintaining the power factor (PF= $\sigma S^2$), is a general strategy to increase $ZT$. Such methods have been demonstrated to be effective in reducing $\kappa_{ph}$, for example, reducing dimensions to bring in size effects on phonon scattering [6,7], creating nanostructures to intensify boundary scatterings [8–10], and changing morphology in a micron scale [10–12]. These strategies aim at designing electron-crystal-phonon-glass structures. In addition, lower-dimensional band structures can augment PF to further boost $ZT$ [13,14].

Organic TE materials provide an alternative blueprint for TE application due to their small $\kappa_{ph}$ [15–19]. Compared with inorganic TE materials, organic TE materials have advantages, for example, light weight, high flexibility, and low cost. Thermal conductivity of organic materials is generally below 1 Wm$^{-1}$K$^{-1}$. Thus conducting polymers, for example, PEDOT and P3HT, are considered as materials with high TE potential [20,21]. Their TE performance can be improved through doping or optimizing molecular morphology [22,23]. The $ZT$ of PEDOT:PSS is up to 0.42 at 300K from experimental measurements [20]. Another type of promising TE candidates is small-molecular semiconductors that are easier to be purified and crystallized than polymers. In small-molecular semiconductors, strong intermolecular electronic couplings provide main paths for their salient charge transport [24]. The mobility of single-crystal rubrene [25] can be up to 43 cm$^2$V$^{-1}$s$^{-1}$. Pentacene thin films doped with iodine [17] have $\sigma$ of

110 S/cm. This excellent electronic transport and poor thermal transport make small-molecular semiconductors competitive in TE applications. More interestingly, the charge transport in three-dimensional (3D) structures of small-molecular semiconductors is strongly anisotropic, where an inter-planar charge transport dominates. Thus TE performance along a non-bonds direction is worthy of exploration. Recent investigations have shown that $C_n$-BTBTs and bis(dithienothiophene) (BDT) have high $ZT$ along the inter-planar direction due to their low-dimensional charge transport [26,27]. Overall, low $\kappa_{ph}$ and the special electronic properties make stacking organic semiconductors become promising candidates in TE applications.

Polymeric carbon nitrides (PCN) have recently attracted a considerable interest as crystalline all-organic wide-bandgap semiconductors in solar water-splitting [28] and photovoltaics [29]. Tyborski *et al* experimentally found that their samples of PCN mainly consist of flat and buckled A-A-stacked structures with a stacking distance of 3.28 Å [30]. The flat structure is shown in Figure 1(a) and Figure 1(b). Merschjann *et al* demonstrated that PCN exhibit a one-dimensional (1D) inter-planar charge transport, which means that the charge transport in PCN is essentially confined to move along one dimension, the *c* direction of the crystal structure [31]. This low-dimensional charge transport property is similar to BDT, where the 1D band structure of BDT results from a $p_z$-orbital overlap and is considered to be responsible for its high $ZT$ [27]. Similarly, a previous model of π-stacked molecular junctions has been reported to possess high TE coefficients [32,33]. For this reason, we are intended to investigate cross-plane TE performance of PCN under the influence of the $p_z$-orbital overlap. Another reason is the low-dimensional advantage of their charge transport. Common small-molecular semiconductors possess herringbone structures that result in two-dimensional (2D) charge transport [34]. The A-A stacked PCN has an 1D charge transport that is preferable in TE applications [35].

In this work, we calculate TE coefficients of A-A stacked PCN along the *c* direction using density function theory (DFT) combined with Boltzmann transport equation (BTE). Its lattice thermal conductivity is calculated by molecular dynamics simulations. To accurately predict the lattice thermal conductivity of A-A stacked PCN along the *c* direction, we consider van der Waals (vdW) interaction and electrostatic forces to describe the overlap of $p_z$ orbitals.

**Methods**

The lattice and electronic structure of PCN are calculated using the Vienna Ab Initio Simulation Package (VASP) [36,37]. The Perdew−Burke−Ernzerhof (PBE) version of the generalized gradient approximation (GGA) is used for the exchange−correlation functional [38]. The DFT-D2 method of Grimme is used to take into account the van der Waals (vdW) interactions among different layers [39]. The cutoff energy of the plane wave expansion is set to 600 eV. The reciprocal space is sampled by a 9×9×15 Monkhorst-Pack k-mesh. A finer mesh of 11×11×31 is used for density of state (DOS). Other computational details are provided in Supporting Information (SI).

The electronic transport coefficients are calculated with BoltzTraP [40] that is based on the semi-classical Boltzmann transport theory. Different from an assumption of constant relaxation time, we calculate electronic relaxation time via a deformation potential approximation (DPA) that is proposed by Bardeen and Shockley [22,26,41,42]. The longitudinal acoustic phonon mode is considered in electron-phonon (e-ph) coupling along the direction of electronic flow. As PCN has 1D electronic transport properties, we use a 1D formula for relaxation time that is expressed as follows.

$$\frac{1}{\tau(\vec{k})} = \frac{k_B T E^2}{\hbar C |v_k|}$$

In the formula, $k_B$ is the Boltzmann constant; T is the absolute temperature; E is the deformation potential constant; $\hbar$ is the reduced Planck constant; C is the elastic constant along the *c* direction; $v_k$ is the group velocity. Details of formula derivations and transport coefficient calculations are provided in SI.

The thermal conductivity of PCN is calculated from equilibrium molecular dynamics (EMD) using the Green-Kubo formula,

$$\kappa = \frac{1}{3k_B T^2 V} \int_0^\infty <\vec{J}(\tau) \cdot \vec{J}(0)> d\tau$$

where κ, $V$, and $T$ are the thermal conductivity, the volume of simulation cell, and temperature, respectively. $\vec{J}(\tau) \cdot \vec{J}(0)$ is the heat current autocorrelation function (HCACF). The angular bracket denotes ensemble average. In this study, all the MD simulations are performed using LAMMPS packages [43]. For such layered structured PCN, the overlap of $p_z$ orbitals introduces a new electronic transport channel. For thermal transport, AMBER force field [44] is used because it has been shown to successfully reproduce thermal transport of layered structures. Previous investigations demonstrated that the geometry of the overlap of $p_z$ orbitals are controlled by electrostatic forces and vdW interactions [45,46]. To accurately predict the lattice thermal conductivity between interlayers, both are considered. The charge distribution of PCN is computed by an atomic charge calculator, which is based on the Electronegativity Equalization Method (EEM) [47–51]. The periodic boundary condition is used in three directions to simulate the infinite system. However, the calculated thermal conductivity still exhibits the finite size effects related to the variation in the longest phonon wavelength that the system can reproduce. In our simulations, 6×6×10 unit cells (100.35×76.11×7.28 nm$^3$) are used, where the calculated thermal conductivity saturates to a constant. More details about EMD simulations can be found in the SI. All the results reported in our studies are ensemble averaged over 12 independent runs with different initial conditions.

**Results and discussion**

From Figure 2(a), for the conduction band (CB), we observe the band along Γ−Z−Γ is much more dispersive than other directions, as well as the bandwidth. Then we use the 1D tight-binding model to fit the CB (seen in SI), and obtain the hopping energy along the c direction up to 0.224 eV. This value is close to hopping energy of hole in BDT, and thus indicates an equivalent strong electronic coupling along the *c* direction. . Comparatively, the band Γ−X and Γ−Y directions is flat, which indicates a smaller electronic hopping energy and thus smaller in-plane electronic couplings. On the other hand, from the projected band structure and the partial density of states, we can learn that the strong coupling originates from the overlap of vertical $p_z$ orbitals. Meanwhile, the charge density of CB in Figure 2(b) shows that nitrogen atoms at the top corners of triangles do not contribute $p_z$ orbitals to CB. That is, electrons are confined within the *a-b* plane but preferable to move along the *c* direction. This feature is consistent with the conclusion from the reference that PCN has a predominantly inter-planar electronic transport [31]. The predominantly inter-planar, or 1D-like electronic transport property of PCN is advantageous in TE applications [52]. The density of states (DOS) of PCN also show a large asymmetry around the Fermi level. It indicates large Seebeck coefficients of PCN as Seebeck coefficients involve an integral over eigenvalues around the Fermi level.

With the knowledge of the inter-planar electronic transport of PCN, we have explored another two similar types of stacked carbon nitrides, *g*-CN1 and *g*-CN2 [53], lattice structures of which are shown in Figure 1(c)-1(f). We calculate their band structures that are shown in Figure 2(e) and Figure 2(f). The CB and valence band (VB) of both band structures are more dispersive along Γ−M and Γ−K than those along Γ−A, which indicates stronger in-plane electronic couplings thus larger electrical conductivities along in-plane directions, compared with those along the out-of-plane direction. That is to say, the electronic transport along in-plane directions dominates in *g*-CN1 and *g*-CN2. Therefore, we focus on PCN in following TE performance calculations.

We calculate TE coefficients of PCN along the *c* direction, using the band model and the Boltzmann transport equation. To validate our band model for charge transport in organic materials, we calculate hopping rates of the CB of PCN, and the reorganization energy of melem monomers that consist of PCN. Details are provided in SI. The hopping rates of the CB are comparable to the reorganization energy, so the band model is acceptable [24,54]. Figure 3(a)-3(d) show the TE coefficients as a function of doping concentration. Seebeck coefficients under 300 K decrease from 50 to 500 μV/K when the carrier concentration increases from $10^{19}$ to $10^{21}$ cm$^{-3}$. These Seebeck values are comparable to inorganic materials, for example, Silicon and single-layer $MoS_2$ that have *ZT* maximum in the same concentration range and have Seebeck up to 300 μV/K under 300 K [8,55].

The electrical conductivity of PCN increases when the carrier concentration increases. In the range between $10^{19}$ and $10^{21}$ cm$^{-3}$, the cross-plane electrical conductivity in PCN has an order of magnitude of $10^5$ S/m. We calculated the intrinsic mobility of PCN that is about 8.55 cm$^2$V$^{-1}$s$^{-1}$ for electrons under 300K, which give a reasonable explanation for its high electrical conductivity. This mobility value is high for organic materials but not restricted to PCN. As Troisi suggests [56], if the reorganization energy is similar to or smaller than rubrene (the inner part is 159 meV [57]), and the vdW interaction between atoms is involved in the frontier orbitals of neighboring molecules, essentially any highly purified crystalline molecular material may have a sufficiently large mobility in the 1–50 cm$^2$V$^{-1}$s$^{-1}$ range. It indicates that π-stacking organic materials with low reorganization energy values have a favorable cross-plane charge transport.

In organic materials, polaron theories and band theories can be both used to describe interactions between charges and the structure deformation [56,58]. The band theories, such as polaronic bands and electronic bands can successfully predict the negative mobility dependence on the temperature [59]. In addition, the band theories are able to predict the high mobility that is consistent with experiments. Shuai *et al* adopted the electronic band model to calculate electronic mobility of pentacene and rubrene that agrees well with experimental measurements [60–62]. In terms of the band theories that are widely used in inorganic semiconductors, electron-phonon couplings are an important contribution to the electronic relaxation time. In our calculations, we have considered the contribution of acoustic phonon modes. Specially, if the polarity of materials is strong, the contribution of polar optical phonon (POP) scattering can be also important, such as GaAs and GaN [63,64]. In these materials, the electronegativity is a key parameter to evaluate the polarity. Back to the carbon nitrides, since the electronegative difference between carbon and nitrogen is larger than that between gallium and arsenic, and we can find that in Figure 2(b), the electron density distribution of PCN is not symmetric, we believe that the polarity of PCN and its POP scattering effects make our calculations overestimate the relaxation time and the electrical conductivity of PCN.

Figure 3(b) shows that the thermal conductivity of PCN, electronic part $\kappa_e$ and lattice part $\kappa_{ph}$. Within the range from $10^{20}$ to $10^{21}$ cm$^{-3}$, $\kappa_e$ is comparable to or even larger than $\kappa_{ph}$. As for phonon transport, the lattice thermal conductivity is 0.77 Wm$^{-1}$K$^{-1}$ at 300 K. To clarify the contribution of the overlap of $p_z$ orbitals to lattice thermal conductivity, other simulations are performed just considering vdW interactions between interlayers. It is found that the lattice thermal conductivity is 0.55 W-m$^{-1}$K$^{-1}$ at 300 K within the range (0.1–1.0 m$^{-1}$K$^{-1}$) [65–69] of the common amorphous polymers due to the weak intermolecular interactions of vdW nature (more details in SI). Therefore, compared with the weak vdW interactions between interlayers, the overlap of $p_z$ orbitals in PCN does noticeably enhance the thermal transport.

With the electronic and phononic properties, we obtain *ZT* profiles of PCN that are shown in Figure 3(c). The optimal *ZT* values at the room temperature are 0.52 with an n-type carrier concentration of $1.23 \times 10^{20}$ cm$^{-3}$ and 0.28 with a p-type carrier concentration of $5.45 \times 10^{20}$ cm$^{-3}$. At the optimal carrier concentration of electrons, the

PF is 17.92 μWcm$^{-1}$K$^{-2}$ and the total thermal conductivity is 1.03 Wm$^{-1}$K$^{-1}$. The p-type thermoelectric performance is overestimated, as the valence band does not exhibit a large bandwidth along the Γ−Z direction. The hole hopping energy along the *c* direction are much smaller than the electronic one, indicating that the band model may lose its validity and overestimate the conductivity of holes. PCN can be doped as n-type materials and contribute to the n-type TE group. Figure 3(d) shows maximum *ZT* values from 100 to 400 K. The *ZT* of PCN has the same order of magnitude as typical herringbone stacked molecules. The calculated optimal *ZT* of pentacene and rubrene are 1.8 and 0.6 at room temperature, respectively [60]. Herringbone stacked BDT is predicted to have peculiar 1D band structure and an optimal *ZT* of 1.48 at room temperature [27].

Due to a capability for TE conversion along the cross-plane direction, PCN can be distinguished from layered materials, specifically inorganics materials such as, graphene, MoS$_2$ and black phosphorus, and thus regarded as a complementary material to 2D TE materials. This feature does not only exist in PCN, as layered inorganic heterostructures also exhibit high *ZT* along the perpendicular directions. Esfarjani *et al* studied TE performance of vdW graphene/phosphorene/graphene heterostructures with the predicted *ZT* of 0.13 under 600K [70]. Similarly, graphene/MoS$_2$ vdW heterostructures are calculated to have *ZT* as large as 2.8 at room temperature [71]. The phonon transport along the cross-plane direction is highly suppressed in both structures. Compared with these nanoscale heterostructures, PCN is a bulk material and should be easier to prepare in experiments and to scale up.

Before closing, we combine PCN, *g*-CN1, *g*-CN2 and A-A stacked graphite, to search for the possible lattice structure that materials can have to obtain high cross-plane *ZT*. The cross-plane direction corresponds to the small lattice thermal conductivity because of no covalent bonds and suppressed lattice vibrations, so strong couplings of electrons or holes along the cross-plane direction are preferable for high *ZT* values. In organic materials, the layer distance and the stacking motif are considered to affect the couplings [72]. For *g*-CN1 and *g*-CN2, they have nearly the same layer distances as PCN, but their band structures exhibit weak electronic couplings between neighboring layers. Our calculated charge density of *g*-CN1 and *g*-CN2 shows a consistent conclusion in Figure 4. We found that $p_z$ orbitals of the CB in Figure 4(a) and 4(c) indicate the existence of the overlap of $p_z$ orbitals. But due to their A-B stacked structures, the coupling is much weaker than PCN as the coupling distance is nearly twice as that in PCN. The $p_z$ orbitals of the VB in Figure 4(b) and 4(d) lie horizontally, showing that holes cannot transfer effectively along the cross-plane direction. So A-A stacked motif or a small coupling distance will be essential for strong inter-planar electronic couplings and the excellent cross-plane TE performance.

Besides, the in-plane electronic confinement is necessary. For A-A stacked graphite, the stacking motif and the electronic coupling distance are nearly the same as PCN. However, electrons and holes are more likely to move within the plane of graphite, because $p_z$ orbitals in graphite form delocalized π bonds in plane. In contrast, the in-plane electronic confinement is caused in PCN because the nitrogen atoms connecting heptazine-like units contribute no $p_z$ orbitals of the CB. Thus, the $p_z$ orbitals of PCN are discontinuously distributed and π bonds are localized along in-plane directions. In such case, charges are more likely to transport from one layer to another and form the 1D-like charge transport. Overall, we conclude that small electronic coupling distances along cross-plane direction and the in-plane confinement effect are responsible for promising cross-plane TE performance.

**Conclusions**

We investigate TE performance of A-A stacked PCN using first-principles calculations and classical molecular dynamics simulations. We find that PCN has a high *ZT* up to 0.52 at 300 K along the *c* direction, which can contribute to n-type TE groups. The overlap of $p_z$ orbitals and the in-plane electronic confinement induce a 1D charge transport and large cross-plane power factors that are comparable to silicon and single-layer $MoS_2$. The lattice thermal conductivity contributed by the overlap of $p_z$ orbitals is considered and found to enhance the thermal transport. By comparing three carbon nitrides and A-A stacked graphite, we find that small electronic coupling distances along cross-plane direction and the in-plane confinement effect induce the 1D charge transport and TE advantages. This study can benefit the pursuing of a high cross-plane *ZT* in other layered materials.

**Acknowledgments**

The work were supported by the National Natural Science Foundation of China, Grant No. 51576076 (NY), 51711540031 (NY), 61371015 (JL), Hubei Provincial Natural Science Foundation of China (2017CFA046), and NSFC-JSPS cooperative projects, Grant No. 51711540031 (NY and JS). The authors are grateful to Dr. Jinyang Xi, Prof. Zhigang Shuai and Prof. Gang Zhang for useful discussions. The authors thank the National Supercomputing Center in Tianjin (NSCC-TJ) and China Scientific Computing Grid (ScGrid) for providing assistance in computations.

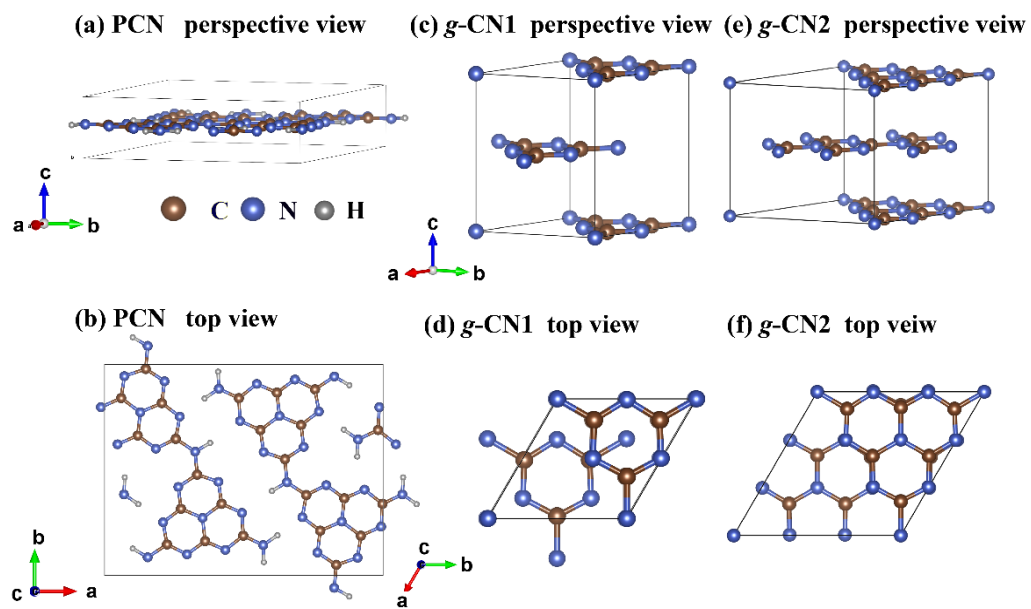

**Figure 1.** Lattice structures of PCN ((a) and (b)), *g*-CN1 (*g*-C$_3$N$_4$) ((c) and (d)) and *g*-CN2 (tri-*g*-C$_3$N$_4$) ((e) and (f)).

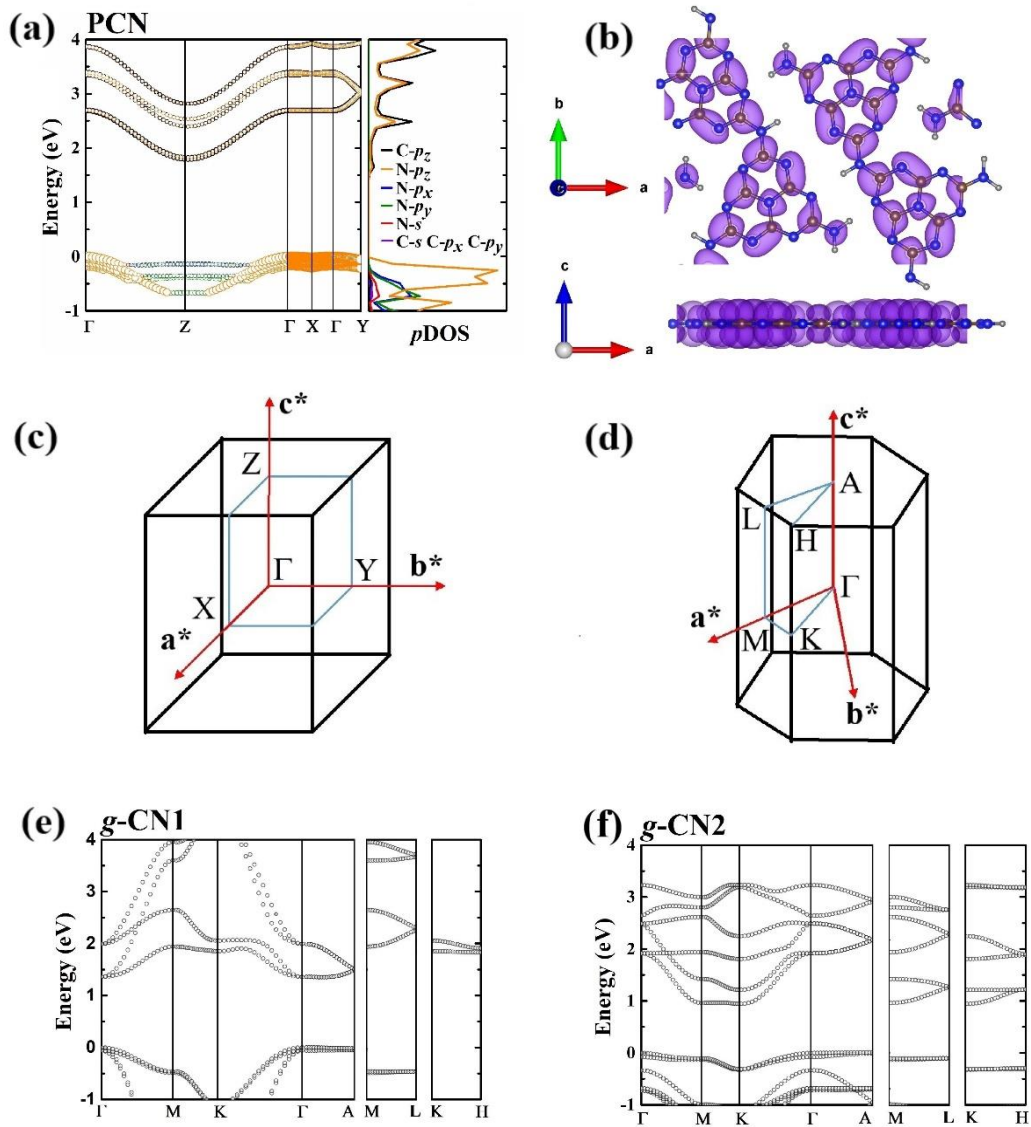

**Figure 2.** (a) Projected band structure and $p$DOS of PCN. (b) Charge density of conduction band of PCN. The purple space shows the charge density. The isosurfaces value is set as $2\times10^{-7}$ e/Bohr$^3$ for positive and negative cases. (c) The Brillouin zone of the orthorhombic lattice and high-symmetric paths for the PCN band structure. (d) The Brillouin zone of the hexagonal lattice and high-symmetric paths for band structures of $g$-CN1 and $g$-CN2. The a* b* and c* are reciprocal lattice vectors. (e) The band structure of $g$-CN1. (f) The band structure of $g$-CN2.

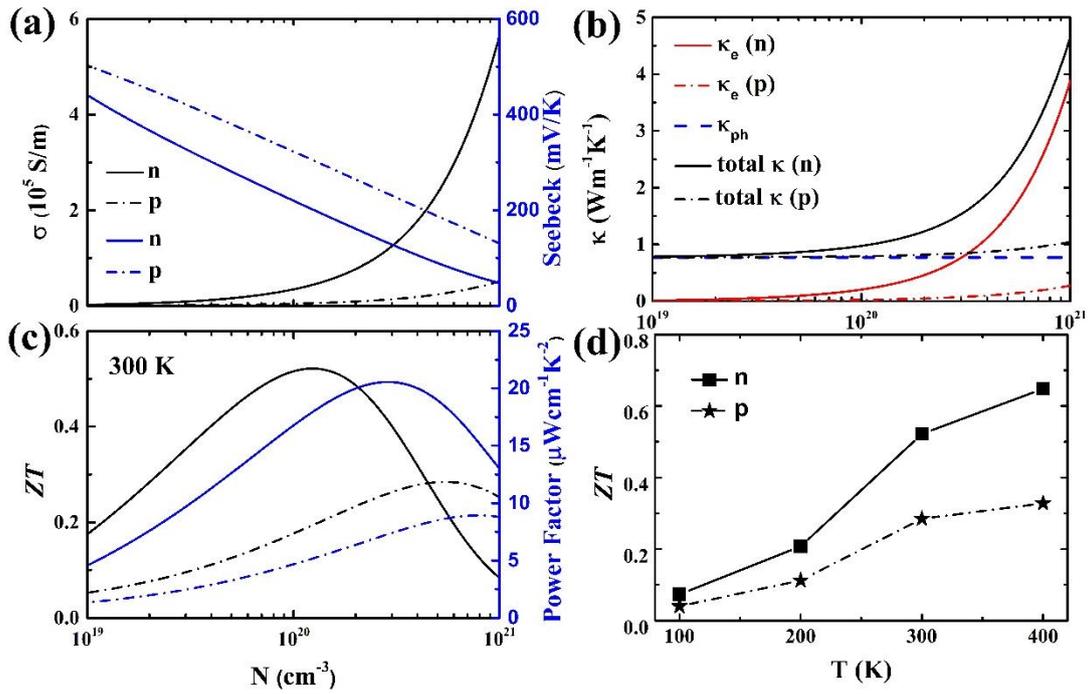

**Figure 3.** (a) (b) (c) TE coefficients of PCN versus n-type (n) and p-type (p) carrier concentrations under 300 K: (a) The electrical conductivity (σ) and Seebeck coefficients; (b) electronic ($κ_e$), lattice ($κ_{ph}$) and total (κ) thermal conductivity; (c) *ZT* and the power factor; (d) The *ZT* maximum dependence of the temperature from 100 to 400 K.

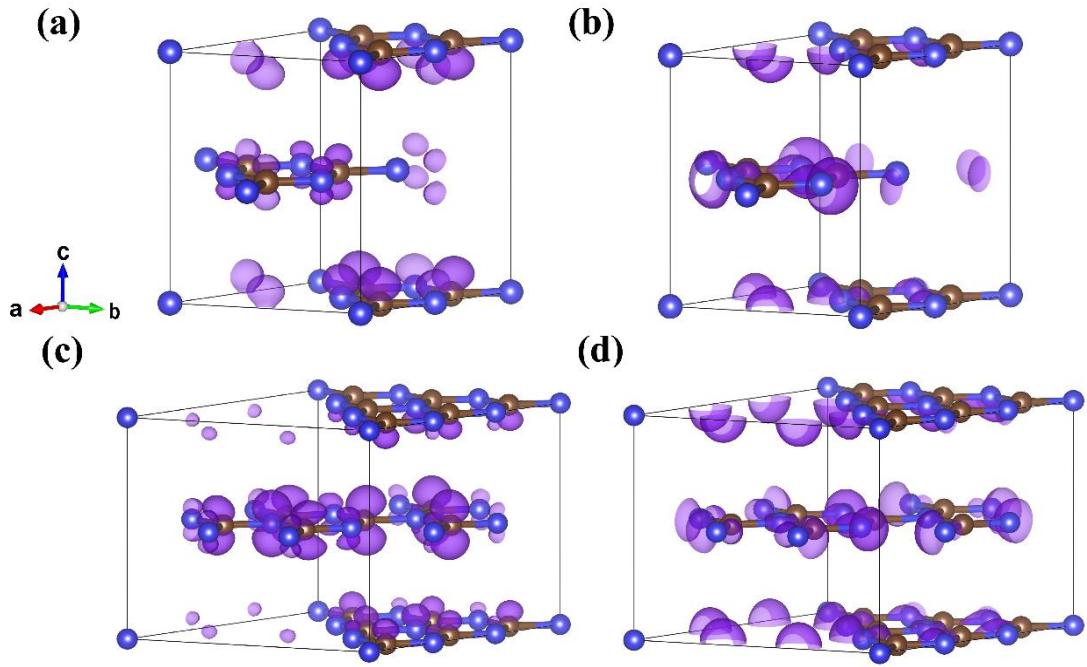

**Figure 4.** The charge density of CB and VB of *g*-CN1 and *g*-CN2: (a) CB of *g*-CN1; (b) VB of *g*-CN1; (c) CB of *g*-CN2; (d) VB of *g*-CN2. For (a) and (b), the isosurfaces value is set to be 0.01 e/Bohr$^3$ for positive and negative cases. For (c) and (d), the isosurfaces value is set to be 0.005 e/Bohr$^3$.